# Energy-Efficient Fracturing Based on Stress-Coupled Perforation

CHENG Cheng[1, 2]

(1. *State Key Laboratory of Oil and Gas Reservior Geology and Exploitation, Chengdu University of Technology, Chengdu, China;*
2. *Department of Chemical and Petroleum Engineering, University of Pittsburgh, Pittsburgh, PA, USA*)

**Abstract:** Hydraulic fracturing is one of the key technologies for reservoir stimulation in low-permeability/unconventional oil and gas fields. In response to the high energy consumption and greenhouse gas emissions caused by extreme flow-limiting perforation in hydraulic fracturing, stress-coupled perforation technology has been proposed to promote low-energy consumption and high efficiency in artificial fracturing. Based on the stress distribution in perforation hole and fracture propagation mechanism, a mathematical model for fracture propagation was established based on linear elastic fracture mechanics theory. Taking into account rock mechanical parameters, tensile effects at the crack tip, stress on both sides of the main crack and fracturing parameters, the real-time stress distribution and fracturing energy consumption were calculated using Monte Carlo random method and Newton's iterative method. With the unit fracture area energy consumption, total fracture area, and fracture uniformity as objective functions, the number and diameter of perforation holes were fully coupled and optimized. The developed simulator balances the calculation efficiency and accuracy of multiple fracture propagation. The study shows that under the condition of the same reservoir fracture volume, stress-coupled perforation technology will reduce energy consumption by 37%. This technology has been successfully applied to the reservoir stimulation of Fan Ye 1 well horizontal well in Jiyang Depression, improving the energy utilization efficiency of hydraulic fracturing and reducing carbon emissions.

**Key words:** multi-stage fracturing; stress-coupled perforation; dynamic stress distribution; multi-fracture growth; energy consumption

## 0 Introduction

Hydraulic fracturing (HF) is playing an increasingly important role in the production of low-permeability/unconventional oil and gas resources. Horizontal well fracturing involves the creation of fractures by injecting fracturing fluid under high pressure into the reservoir through casing from 3 to 6 perforation clusters simultaneously, and then the proppant is used to keep the fractures from closing, allowing oil and gas to flow along the fractures under formation pressure. However, 20% to 40% of perforation clusters make no contribution to production[1]. Factors leading to this uneven production from perforation cluster groups not only include the in-situ stress along the wellbore, the heterogeneity of reservoir physical and mechanical properties[2,3], but also involve "stress shadowing", that is, the suppression of fracture propagation due to the compressive stress applied by hydraulic fracturing[4-7]. This uneven fracture propagation is not conducive to the full utilization of the injected fluid, leading to a reduction in the fractured area under the same injection volume, and since production is positively correlated with fracture area, this in turn reduces productivity.

The perforation process has a significant impact on the effectiveness of hydraulic fracturing. In recent years, the Extreme Limited Entry (EXL) perforation has been widely used. The principle of this technique is to dominate the global pressure in the wellbore by creating fewer or smaller diameter holes in the perforation clusters, controlling fracture propagation through pressure drop. When the dynamic distribution of stress and fracture propagation rules are not mastered at the actual construction site, blindly using this technique to promote even fracture propagation can lead to the pressure drop at the perforation hole several times the fracturing pressure [8-10]. This requires a significant increase in pumping power or pumping pressure to maintain a sufficient injection rate for fracture propagation. Although this technique reduces the uncertainty of fracture development, it also induces higher cost. According to data from the U.S. Energy Information Administration, fracturing pumping can cost 41% of total investment[11]. Therefore, the use of the EXL perforating technique not only requires high-power pumps and higher fuel consumption, leading to serious environmental pollution [12] and waste of resources [13,14], but also risks of water resource management, negatively impacting economic benefits. In addition, it also increases the risk of seismic activity [15]. These negative impacts are essentially all related to the increased energy consumption of hydraulic fracturing [16-18].

Hydraulic fracturing technology needs to reduce its environmental impact by lowering energy consumption. Furthermore, environmental and social impact are not the only concern, maximizing the return on investment is equally important. In fact, the greenhouse gas (GHG) emissions per unit of energy related to well completion (i.e., kg CO2eq/MWh) can be reduced by lowering the energy consumption per unit of production[19,20]. Therefore, we emphasize reducing energy consumption in the shale oil and gas extraction process from two aspects: Promoting fracture development and improving oil and gas productivity while reducing energy consumption, that is, reducing the energy consumption per unit of production.

In order to achieve the goals of fracture expansion and energy consumption reduction, a stress-coupled perforation technique (SCP) considering dynamic stress interference between wells, stages, and clusters is proposed. Based on the mechanical mechanism of fracture expansion in elastic media, this technique balances the coupling relationship between geostress and elastic stress. While maintaining oil and gas productivity, it achieves up to a 40% reduction in energy consumption, thus reducing the cost and carbon emissions of artificial fracturing.

This paper first introduces the mathematical model of fracture expansion based on the theory of linear elastic fracture mechanics. Based on this, the Monte Carlo method and Newton's iteration method are preferentially selected to randomly select perforation parameters and calculate the dynamic stress distribution and fracture development scale during the fracturing process. The energy consumed per unit fracturing area is taken as the objective function to optimize the combination of perforation parameters.

## 1 Mathematical Model

### 1.1 Calculation Method for Dynamic Stress Distribution

Stage fracturing is a dynamic process in which the stress undergoes variation in the surrounding medium of operation wellbore. The stress-coupled perforation technique is the optimal combination of perforation hole count and hole diameter, based on the accurate coupling between the prediction of dynamic stress distribution and the perforation parameters.

To precisely calculate its numerical value, each fracture corresponding to a perforation cluster is first matrixed in the form of $F_{i,j,k}$. Here, $i$ refers to the $i$-th well in the well group, $j$ refers to the $j$-th stage in the well respectively, and $k$ denotes the $k$-th fracture within a stage, with $k$=1 near the toe and $k$=5 near the heel (Figure 1). Once each fracture is marked, the normal stress at any point in space can be calculated according to the elastic stress equation. Assuming an arbitrary point C in the reservoir with a horizontal coordinate of $x$, a vertical coordinate of $y$, and a height coordinate of $z$, the method for calculating the normal stress in point C due to deformation of fracture $F_{i,j,k}$ is illustrated as follows: The fracture $F_{i,j,k}$ corresponds to the spatial position A of the perforation cluster ($x_{i,j,k}$, $y_{i,j,k}$, $z_{i,j,k}$), and the fracture $F_{i,j,k+1}$ corresponds to the spatial position B of the perforation cluster ($x_{i,j,k+1}$, $y_{i,j,k+1}$, $z_{i,j,k+1}$). We can get the vectors $\overrightarrow{AC} = (x - x_{i,j,k}, y - y_{i,j,k}, z - z_{i,j,k})$ and $\overrightarrow{AB} = (x_{i,j,k-1} - x_{i,j,k}, y_{i,j,k-1} - y_{i,j,k}, z_{i,j,k-1} - z_{i,j,k})$. The projection length of vector $\overrightarrow{AC}$ on vector $\overrightarrow{AB}$ is given by $(\overrightarrow{AC} \cdot \overrightarrow{AB}) / |\overrightarrow{AB}|$, where $|\overrightarrow{AB}|$ denotes the modulus of vector $\overrightarrow{AB}$, that is, the length from point A to point B. Then, using the Pythagorean theorem, the vertical distance $D_c$ from any point C to the plane perpendicular to the well trajectory passing through the perforation cluster ($x_{i,j,k}$, $y_{i,j,k}$, $z_{i,j,k}$) can be obtaind through:

$$\mathrm{D}_{C,(i,j,k)} = \sqrt{|\overrightarrow{AC}|^2 - \left[(\overrightarrow{AC} \cdot \overrightarrow{AB}) / |\overrightarrow{AB}|\right]^2} \quad (1)$$

$$\mathrm{L}_{C,(i,j,k)} = \sqrt{|\overrightarrow{AC}|^2 - \left[\mathrm{D}_{C,(i,j,k)}\right]^2} \quad (2)$$

记：

$$\zeta_{C,(i,j,k)} = \frac{\mathrm{D}_{C,(i,j,k)}}{\mathrm{R}_{i,j,k}} \quad \rho_{C,(i,j,k)} = \frac{L_{C,(i,j,k)}}{R_{i,j,k}} \quad (3)$$

Here, $|\overrightarrow{AC}|$represents the modulus of vector$|\overrightarrow{AC}|$, which is the distance from point C to point A. The distance from the projection point of any point C in space on the fracture surface $F_{i,j,k}$ to the position ($x_{i,j,k}$, $y_{i,j,k}$, $z_{i,j,k}$) of the corresponding perforation cluster is calculated as:

$$\mathrm{L}_{C,(i,j,k)} = \sqrt{|\overrightarrow{AC}|^2 - \left[\mathrm{D}_{C,(i,j,k)}\right]^2} \quad (2)$$

Let:

$$\zeta_{C,(i,j,k)} = \frac{\mathrm{D}_{C,(i,j,k)}}{\mathrm{R}_{i,j,k}} \quad \rho_{C,(i,j,k)} = \frac{L_{C,(i,j,k)}}{R_{i,j,k}} \quad (3)$$

$\zeta_{C,(i,j,k)}$ is the ratio of the projection vertical distance $D_{C,(i,j,k)}$ to the radius of the fracture $R_{i,j,k}$. As the fracture expands, $R_{i,j,k}$ increases, and the value of $\zeta_{C,(i,j,k)}$ decreases. $\rho_{C,(i,j,k)}$ is the ratio of the distance $L_{C,(i,j,k)}$ between the projection point of point C on the fracture surface and the center of the fracture $F_{i,j,k}$ to the fracture radius $R_{i,j,k}$.

$$P_{i,j,k} = \frac{3E'V_{i,j,k}}{16{R_{i,j,k}}^3} \quad (5)$$

$P_{i,j,k}$ is the modified uniform internal net pressure generated by the uniformly pressurized ellipsoidal fracture. Considering that the full elastic solution of each non-uniform and transient internal fracture pressure is the main computational bottleneck. To calculate quickly, Cheng and Bunger proposed an approximation in 2016 in which the non-uniform pressure is replaced by uniform pressure, and at each time step, this uniform pressure is chosen to produce a fracture with the same volume as the actual fracture opened by non-uniform internal pressure[21], that is:

$$V_{i,j,k} = 2\pi \int_0^{R_{i,j,k}(t)} w_{i,j,k}(r,t) r dr \quad (6)$$

Based on the above parameters, according to equations (1-6), the normal stress component applied by the fracture $F_{i,j,k}$ to any point C in space is approximately

given by:

$$\sigma_{C,(i,j,k)} = \frac{2P_{i,j,k}}{\pi}\left\{\begin{array}{c} \delta_{C,(i,j,k)}^{-\frac{1}{2}} \cos\frac{1}{2}\varphi_{C,(i,j,k)} - tan^{-1} \dfrac{\left[\begin{array}{c}\delta_{C,(i,j,k)}^{\frac{1}{2}} \sin\frac{1}{2}\varphi_{C,(i,j,k)} \\ +\tau_{C,(i,j,k)} \sin\theta_{C,(i,j,k)}\end{array}\right]}{\left[\begin{array}{c}\delta_{C,(i,j,k)}^{\frac{1}{2}} \cos\frac{1}{2}\varphi_{C,(i,j,k)} \\ +\tau_{C,(i,j,k)} \cos\theta_{C,(i,j,k)}\end{array}\right]} \\ +\zeta_{C,(i,j,k)}\delta_{C,(i,j,k)}^{-\frac{3}{2}} \\ \cos\left(\frac{3}{2}\varphi_{C,(i,j,k)} - \theta_{C,(i,j,k)}\right) \\ -\zeta_{C,(i,j,k)}\delta_{C,(i,j,k)}^{-\frac{1}{2}} \sin\frac{1}{2}\varphi_{C,(i,j,k)} \end{array}\right\} \quad (4)$$

Here, other variables are formed by the fracture length $R_{i,j,k}$ of fracture $F_{i,j,k}$ and Equations (7-10).

$$\tau_{C,(i,j,k)} = \left(1 + \zeta_{C,(i,j,k)}^2\right)^{\frac{1}{2}} \quad (7)$$

$$\theta_{C,(i,j,k)} = arctan\left(\frac{1}{\zeta_{C,(i,j,k)}}\right) \quad (8)$$

$$\delta_{C,(i,j,k)} = \left\{\begin{array}{c}\left[\left(\rho_{C,(i,j,k)}\frac{R_{i,j,k}}{L_{C,(i,j,k)}}\right)^2 + \zeta_{C,(i,j,k)}^2 - 1\right]^2 \\ +4\zeta_{C,(i,j,k)}^2\end{array}\right\}^{\frac{1}{2}} \quad (9)$$

$$\varphi_{C,(i,j,k)} = arccot \left\{\begin{array}{c}\left[\left(\rho_{C,(i,j,k)}\frac{R_{i,j,k}}{L_{C,(i,j,k)}}\right)^2 + \zeta_{C,(i,j,k)}^2 - 1\right] \\ /2\zeta_{C,(i,j,k)}\end{array}\right\} \quad (10)$$

Considering the deformation caused by fracturing during the hydraulic fracturing process of a well factory, which exists between wells, stages and clusters, these deformations subsequently change the stress distribution in the formation. As stage fracturing is a dynamic process, the magnitude of stress at any point C in space is a result of stress accumulation, as shown below:

$$\sigma_C = \sum_i^N \sum_j^{n_i} \sum_k^{m_j} \sigma(\zeta_{C,(i,j,k)}, \rho_{C,(i,j,k)}, t) \quad (11)$$

Where N is the number of wells in the well group that have undergone hydraulic fracturing, $n_i$ is the number of stages that have been fractured in the $i$-th well, and $m_j$ is the total number of fractures that are simultaneously fractured in the $j$-th stage. To calculate the stress distribution of the surrounding rock of the operation well group, just replace the coordinates of point C in above Equation (11) with the coordinates of the fracture surface, and then the magnitude of the normal stress applied on any fracture can be obtained.

## 1.2 Calculation Method of Fracture Propagation

During the process of stage fracturing, multiple perforation clusters need to be set up initially in the operation part of the wells. Then high-pressure fracturing fluid is injected simultaneously into each perforation cluster. Based on the propagation conditions of Linear Elastic Fracture Mechanics (LEFM), when the fracture criterion given by Rice $K_I = K_{Ic}$ (1968) is met, the reservoir will be fractured in the direction perpendicular to the minimum in-situ confining stress $\sigma_{min}$, thus creating a fracture. $K_I$ represents the stress intensity factor of Mode I (opening mode), and $K_{IC}$ represents the fracture toughness of Mode I fracture[22].

Fractures must overcome the stress imposed by the surrounding rock in order to propagate. The stress $\sigma_C$ that the fracture surface suffering is calculated by Equation (4). Using the fracture propagation energy calculation formula, we can determine the energy $W_I$ consumed by the fracture in overcoming the stress during the propagation process. Taking the fracture $F_{i,j,k+1}$ as an example, the work done to overcome the compressive stress from fracture $F_{i,j,k}$ in its expansion is calculated as:

$$\dot{W}_{I(i,j,k+1)} = -2\pi \times \left(\begin{array}{c}\int_{R_w}^{min(R_{i,j,k.}\ R_{i,j,k+1})} \sigma_{i,j,k}\frac{\partial w_{i,j,k+1}}{\partial t} rdr \\ +\sigma_{i,j,k}\frac{dR_{i,j,k}}{dt}R_{i,j,k}w_{i,j,k+1}\left(\frac{R_{i,j,k}}{R_{i,j,k+1}}\right)\end{array}\right) \quad (12)$$

$$P_{perf(i,j,k)} = \left(a\rho/n_{i,j,k}^2 D_{p(i,j,k)}^4 C^2\right)Q_{i,j,k}(t)^3 \quad (17)$$

In addition to overcoming the stress exerted by adjacent and simultaneously fractured fractures, the fracture also needs to overcome the stress induced by the previously cracked fractures. Its power is calculated by the following formula:

$$\dot{W}_{i,j,k} = Q_{i,j,k} \times \int_{z_0}^{z_t}\int_{y_0}^{y_t}\int_{x_0}^{x_t} \sum_i^N \sum_j^{n_i} \sum_k^{m_j} \sigma_{C,(i,j,k)}(x,y,z,t)\, dxdydz \quad (13)$$

The upper and lower limits of the integral in Equation (13) are determined by the boundaries of the fracture in three-dimensional space. By introducing the rate of work done by the fracture to overcome stress shadow into the power balance on the right side of $\dot{W}_f$[23], the influence of stress distribution on flow rate allocation can be obtained.

$$p_{fo}Q_{i,j,k} = \dot{W}_C + D_f \tag{14}$$

The left side of this equation represents the total input power (product of pressure and injection flow rate) of the $k$-th fracture in the $j$-th stage of the $i$-th well. The first term on the right side, $\dot{W}_c$ represents the power consumed by the fluid applied on the solid, mainly including: the rate of work done by the compressive stress induced by fracture $i$ on other fractures $W_I$; the strain energy rate generated by rock deformation $U$; the work done by the fracture to overcome the minimum in-situ stress $\sigma_{min} * Q_{i,j,k}$; the power consumed by the fracture to overcome the stress of already fractured fractures $\dot{W}_{i,j,k}$ ; and the energy dissipation rate related to rock breakage $F_C$:

$$\dot{W}_c = W_I + U + \sigma_{min}Q_{i,j,k} + \dot{W}_{i,j,k} + F_C \tag{15}$$

The second term on the right side of Equation (14) represents the input power consumed by fluid flow, which are the energy dissipation rate related to viscous fluid flow, $F_f$, the energy loss rate related to fluid leak-off into surrounding rock $D_L$, and the power loss of fracturing fluid flowing through perforation holes into formation $P_{perf}$:

$$D_f = F_f + D_L + P_{perf} \tag{16}$$

For detailed explanations of these energy loss rate terms, please refer to the research results published by Cheng and Bunger in 2016[21]. Further explanation is not provided here. The energy consuming rate related to perforations, $P_{perf}$, is expressed by the formula given by Bunger et al. in 2014[24]:

$$P_{perf(i,j,k)} = \left(a\rho/n_{i,j,k}^2 D_{p(i,j,k)}^4 C^2\right) Q_{i,j,k}(t)^3 \tag{17}$$

$P_{perf(i,j,k)}$ is the energy loss at the entrance of fracture $k$ when the fluid passes through, which needs to analyze the pressure drop when the fluid passes through the cluster of $n_{i,j,k}$ perforation holes[25,26]. $Q_{i,j,k}(t)$ is the volumetric flow rate through the $k$-th perforation cluster at time $t$, which varies with the pumping time $t$. $n_{i,j,k}$ is the number of perforation holes in the $k$-th cluster, usually between 5 and 20. $D_{p(i,j,k)}$ represents the perforation diameter of the $k$-th cluster, generally between 6 and 15 millimeters. In addition, the perforation hole itself has a shape factor C, which is 0.56 before erosion and 0.89 after erosion[25]. C5Frac neglects hole erosion, so C is 0.56. The numerical factor $a$ often takes the value of 0.8106. The density of the fluid injected into the reservoir is $\rho$. The variables in the equation brackets form the ratio of power loss passing through the perforation hole. The energy loss of the fluid at the hole is negatively proportional to the fourth power of the hole diameter and the second power of the number of holes. These slight changes in two perforation parameters will have a huge impact on the energy loss of the fluid passing through the hole, thereby directly affecting the fluid distribution between the perforation clusters under the constraint of power balance (Equation 14). These two parameters can be manually controlled and adjusted during the completion stage of fracturing, therefore, as the main parameters affecting the fracturing effect and energy consumption, the number of perforation holes and the hole diameter in a single cluster need to be optimized.

$$T_{i,j,k}\left(\rho_{i,j,k}, \mu_{i,j,k}^c(t), \psi_{i,j,k}(t)\right) = \left(\frac{E'^2\mu_{i,j,k}^c(t)}{t}\right)^{\frac{1}{3}} \left\{ \begin{array}{c} A\left[\omega - \dfrac{2}{3\left(1-\rho_{i,j,k}\right)^{\frac{1}{3}}}\right] \\ -B\left(ln\dfrac{\rho_{i,j,k}}{2}+1\right) \\ +\psi_{i,j,k}(t) \end{array} \right\} - \sum_{i}^{N,N\neq i}\sum_{j}^{n_i,n_i\neq j}\sum_{k}^{m_j,m_j\neq k} \sigma_{(i,j,k)}\left(\zeta_{(i,j,k)}, \rho_{(i,j,k)}, t\right) \tag{18}$$

$$R_{i,j,k}(t) = \gamma_{i,j,k}(t) \cdot \left(\left(\frac{E't}{\mu_{i,j,k}^c(t)}\right)^{1/3} \int_0^t Q_{i,j,k}(t)dt\right)^{1/3} \tag{19}$$

$$w_{i,j,k}(r,t) = \frac{8R_{i,j,k}}{\pi E'} \cdot \int_{\rho_i}^{1} \frac{s}{\sqrt{s^2-\rho_i{}^2}} \int_0^1 \frac{xT_{i,j,k}\left(\rho_{i,j,k}, \mu_{i,j,k}^c, \psi_{i,j,k}\right)}{\sqrt{1-x^2}} dxds \tag{20}$$

Assume the real-time fluid flow rate $Q_{i,j,k}$ of each perforation cluster in the operating section, relying on the approximate solution of fracture length, opening, and flow pressure. The traction force quantifying the effect of inter-well, inter-stage, and inter-cluster stress received by the fracture $F_{i,j,k}$ is:

The last term of the above equation indicates that the traction force has considered stress interference. The fracture length $R_{i,j,k}$ and fracture opening $w_{i,j,k}$ considering the dynamic stress of well group fracturing are further derived from the traction force:

$$R_{i,j,k}(t) = \gamma_{i,j,k}(t) \cdot \left(\left(\frac{E't}{\mu_{i,j,k}^{c}(t)}\right)^{1/3} \int_0^t Q_{i,j,k}(t)dt\right)^{1/3} \quad (19)$$

$$w_{i,j,k}(r,t) = \frac{8R_{i,j,k}}{\pi E'} \cdot \int_{\rho_i}^{1} \frac{s}{\sqrt{s^2-\rho_i{}^2}} \int_0^1 \frac{xT_{i,j,k}(\rho_{i,j,k}, \mu_{i,j,k}^{c}, \psi_{i,j,k})}{\sqrt{1-x^2}} dxds \quad (20)$$

This leads to the development of fracture under different perforation parameter combinations, thus optimizing the number and diameter of perforation holes is available.

### 1.3 Parameter Optimization Model

The formula proposed by Fisher in 2002 reveals the most direct connection between well productivity $q$ and fracture area $A$ [25]:

$$q \approx kh\Delta pA/\mu\Delta xh \quad (21)$$

The fracture area obtained by model simulation directly reflects the level of productivity because it is positively correlated with the stimulated reservoir volume (SRV)[27]. Considering the extension of simulated fractures driven by fluid, the energy of fractures is mainly dissipated in the form of fluid flow rather than fracture rupture, such that fractures will tend to expand in a circular shape[28,29], so the fracture area uses circular surface area: $A = \sum_{i=1}^{n} R_i{}^2\pi$ . Here $n$ is the number of perforation clusters in a single stage. With the increasing use of extreme limited entry (EXL) in more and more oilfields, the accompanying high energy consumption problem leads to carbon emissions and other negative environmental footprints. Considering this factor, the design goal of stress coupled perforation (SCP) is: while maintaining the fracture area equivalent to EXL, the fracturing energy consumption should be minimized. Therefore, the energy consumption per unit fracture area is used as the optimization target function, and a universal method of perforation parameter optimization is established to meet the needs of single well and well group under any geological conditions. Its formula is as follows:

$$E_f = \frac{\sum_{k=1}^{N} \int_0^T p_{fo(k)}(t) Q_{o(k)}\, dt}{\sum_{k=1}^{N} \sum_{i=1}^{n} R_i{}^2\pi} \quad (21)$$

Where $k$ is the $k$-th stage of a single well, $N$ is the total number of stages, $E_f$ is the energy consumption per unit fracture area, and the unit is Kwh/m$^2$. In addition, in order to solve multi-objective optimization problems, we introduce balanced energy consumption $\Delta E$, total fracture area $\Delta A$, and fracture uniformity $\Delta D$. The expression is as follows:

$$\Delta E = \left[\frac{\left(E_{f(m)} - Max\left(E_{f(m)}\right)\right)}{Max\left(E_{f(m)}\right)}\right]^2 \quad (23)$$

$$\Delta A = \left[\frac{\left(A_{T(m)} - Max(A_{T(m)})\right)}{Max(A_{T(m)})}\right]^2 \quad (24)$$

$$\Delta D = \left[\frac{\left(D_{A(m)} - Max(D_{A(m)})\right)}{Max(D_{A(m)})}\right]^2 \quad (25)$$

In the formula, $m$ represents the total number of test perforation parameter samples, and $D_{A(m)}$ is the standard deviation of fracture area of the $m$-th test.

## 2 Algorithm

Most of the commonly used fracturing simulation software are limited by computation efficiency, so fracturing parameter optimization design adopts a decoupling method, which is, only one parameter is allowed to vary, while other parameters to be optimized are set to fixed values. Once a parameter is optimized, it is fixed, and then other parameters are optimized one by one. However, in the actual process of fracture expansion, multiple parameters interact with each other, and decoupling optimization cannot achieve the expected optimization effect.

Unlike conventional methods, the semi-numerical C5Frac simulation model[21], which combines approximate solutions with energy law, enhancing computation speed while ensuring precision. This enables the coupled optimization of perforation parameters based on dynamic stress distribution, effectively reducing the energy consumption in hydraulic fracturing. The algorithm for parameter optimization in Stress Coupled Perforation (SCP) proceeds as follows:

S1—Set initial parameters such as reservoir rock mechanics, clustering, pumping procedure and so on, including $\sigma_{min}, C_L, E, \vartheta, \mu, K_I, Q_o, T, \Delta t, h_i, Z$ etc., where $\sigma_{min}$ is the minimum in-situ stress; $C_L$ is the leak-off factor, used to describe the sealing ability of the surrounding rock to the injected fluid, affected by rock properties and fluid properties; $E$ is Young's modulus, $\vartheta$ is Poisson's ratio, $\mu$ is fluid viscosity, $K_I$ is rock fracture toughness, $Q_o$ is total injection flow rate, $T$ is total pumping time, $\Delta t$ is time step, $h_i$ is the distance between each perforation cluster in certain stage, and $Z$ is stage length.

S2—Using the Monte Carlo random method, a value is randomly selected from between 5 and 20 each time as the number of perforation holes in each perforation cluster within the stage; the diameter of the hole is randomly selected from a value between 0.006 and 0.015m; then these two randomly selected perforation parameters are combined into a trial perforation parameter pair.

S3—The initial solution specifies $Q_i(t_o) = Q_o/N$, where $t_o$ is a specified initial time (before leakage and interaction become important). If it is not the initial time, $Q_i(t)$equals the flow rate $Q_i(t-\Delta t)$ of the previous time step.

S4—Substitute the initial iterative flow rate $Q_i(t)$ into equations (18-20) to obtain the fracture propagation result.

S5—Based on the fracture growth and stress calculation Equation (12), the real-time stress distribution status is obtained. Calculating the power consumption for the fracture to overcome the stress action of the already fractured fracture through Equation (14). Calculating the power loss of the fluid through the perforation hole according to the trial perforation parameter pair and Equation (17), then substituting with other energy power items into the power balance formula (14).

S6—Applying Newton's iterative method to the power balance to obtain the instantaneous flow rate $Q_i(t)$ of each perforation cluster in this iteration. Substituting the flow rate $Q_i(t)$ obtained from this iteration into the third step S3 of the algorithm; repeating steps S3-S6 until the change of $Q_i(t)$ reaches the acceptable convergence level.

S7—After obtaining $Q_i(t)$, repeat steps S3-S6 until the total pumping time $T$ is obtained. Then use Equation (22) to accumulate the total fracture area within a single stage.

S8—Repeating S3-S7, the fracture growth of all stages in a single well and the corresponding total energy consumption is achieved. Then the energy consumption of unit fracturing area of the perforation parameters pair is obtained through Equation (23).

S9—Repeating S2-S8 until all possible combinations (16*10=160) of perforation number and perforation diameter are simulated. Finally, on the premise that the volume of the injected fluid is equal, comparing the unit fracture area energy consumption $E_f$ of each combination of perforation parameters, and selecting the combination of perforation parameters that reaching the maximum fracturing fracture area with the minimum energy consumption.

The implementation of the above algorithmic process relies on the self-developed semi-numerical simulator C5Frac. This simulator offers sufficient precision while maintaining computational efficiency, enabling tens of thousands of evaluations to be run within a manageable time. This facilitates the coupled optimization of perforation parameters under complex geological conditions, reducing energy consumption per unit fracture area, and subsequently lowering carbon emissions

## 3.Application Experiment

The effectiveness of Extreme Limited Entry (EXL) and Stress Coupled Perforation (SCP) is compared through two application cases. Case one involves a homogeneous reservoir for a single well. The designed spacing between sections is 60m, with five perforation clusters symmetrically distributed, and a cluster spacing of 12.5m; the first perforation cluster at the toe of the well is 5m away from the bridge plug, and the first perforation cluster at the heel of the well is also 5m away from the other bridge plug.

Case two is an actual field trial in Fan Ye 1 well factory, involving four horizontal wells with trajectories extending in roughly the same direction and are arranged non-parallel, the distance between the horizontal wells ranges from 200 to 800 meters, and the vertical depths differ by about 100 meters. The target zone are essentially located in the same reservoir, and the heterogeneity of reservoir parameters such as porosity and permeability around the wellbore is strong, and hence the variability of the leak-off coefficient along the wellbore is large.

Based on the aforementioned perforation parameter optimization method, two strategies were developed. Strategy one: for a single well in a general homogeneous reservoir, globally optimizing a set of stress-coupled perforation parameters for all sections called Global Stress Coupled Perforation (G-SCP). Strategy two: for the well group in the heterogeneous reservoir, optimizing perforation parameters section by section. The optimization of perforation parameters for each section is based on the stress field calculations of the operating stage, named Stage Stress Coupled Perforation (S-SCP). This approach is taken to minimize the heterogeneity between sections as much as possible and to ensure that the fracturing development of each section is similar. According to the number of stages in field and the intensity of reservoir heterogeneity, the computational resources consumed by corresponding scientific computing is large-scale, so the global (G-SCP) and local (S-SCP) stress coupled perforation are used flexibly. While taking into account computational efficiency and reduces computational demand, which is another way to optimize energy consumption,

In both cases, the parameters for Extreme Limited Entry (EXL) are 5 holes per cluster and a hole diameter of 6mm; the selection range for Stress Coupled Perforation (SCP) parameters is 5 to 20 holes per cluster and hole diameters ranging from 6 to 15mm.

## 3.1 Example of Global Stress Coupled Perforation in Homogeneous Reservoir for a Single Well

The representative reservoir parameters and engineering parameters such as Carter leak-off coefficient, reservoir pressure, and fluid injection rate are imported into the semi-numerical model C5Frac to compare the differences between the Global Stress Coupled Perforation (G-SCP) and the Extreme Limited Entry (EXL). The parameters are as follows:

$$C_L = 8.89 * 10^{-6} m/s^{\frac{1}{2}};\quad E = 32.16\,\text{GPa};\quad \nu = 0.23;$$

$$C_{ro} = 1.5 * 10^{9}\text{Pa}^{-1};\quad K_{IC} = 1\,\text{MPa}\cdot\text{m}^{\frac{1}{2}};$$

$$\sigma_{min} = 63.63\,\text{Mpa};\quad R_W = 0.2\,\text{m};\quad \mu = 0.003\text{Pa.s};$$

$$P_r = 50.90\,\text{Mpa};\quad Q_O = 0.15\text{m}^3/s;\quad V_{inj} = 540\text{m}^3$$

where $C_{ro}$ is the rock compressibility coefficient, $R_W$ is the wellbore diameter, $P_r$ is the reservoir pressure, and $V_{inj}$ is the total injection volume of fluid per stage. The stage length is set as 60m that commonly used , with 5 perforation clusters distributed symmetrically about the center, with a spacing of 12.5m. The distance between the perforation clusters at both ends and the plug is 5m. The lateral length is 2160m, with a total of 36 stages. Monte Carlo method is used to random combine the number of perforations and the diameter of the perforations as candidate ready for numerical experiments with the known geological and engineering parameters. After fully coupled computation, and the design parameters of the Global Stress Coupled Perforation (G-SCP) are finally obtained shown in Table 1.

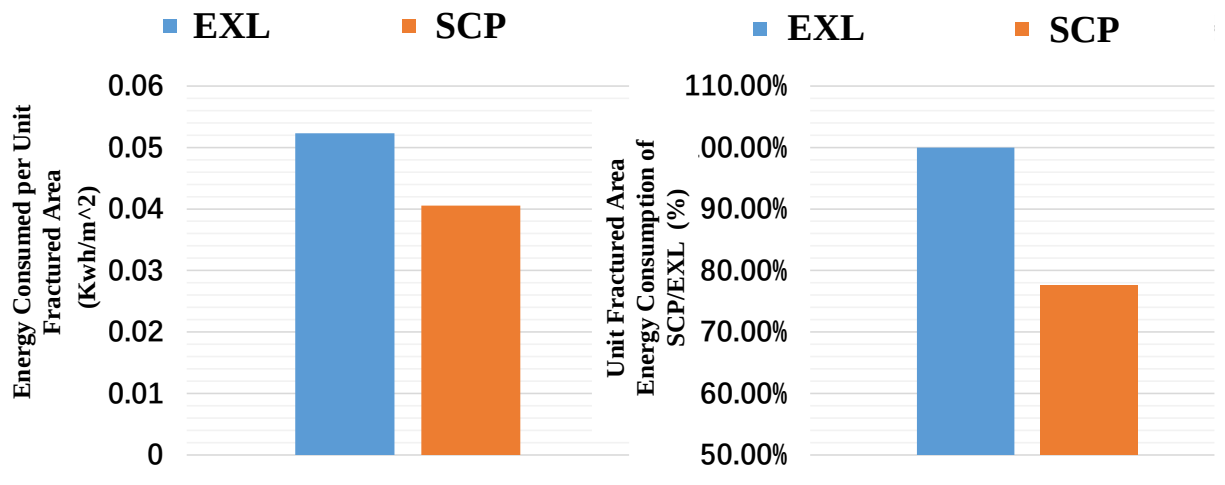

Figure 4 Comparison of Energy Consumption per Unit Fractured Area

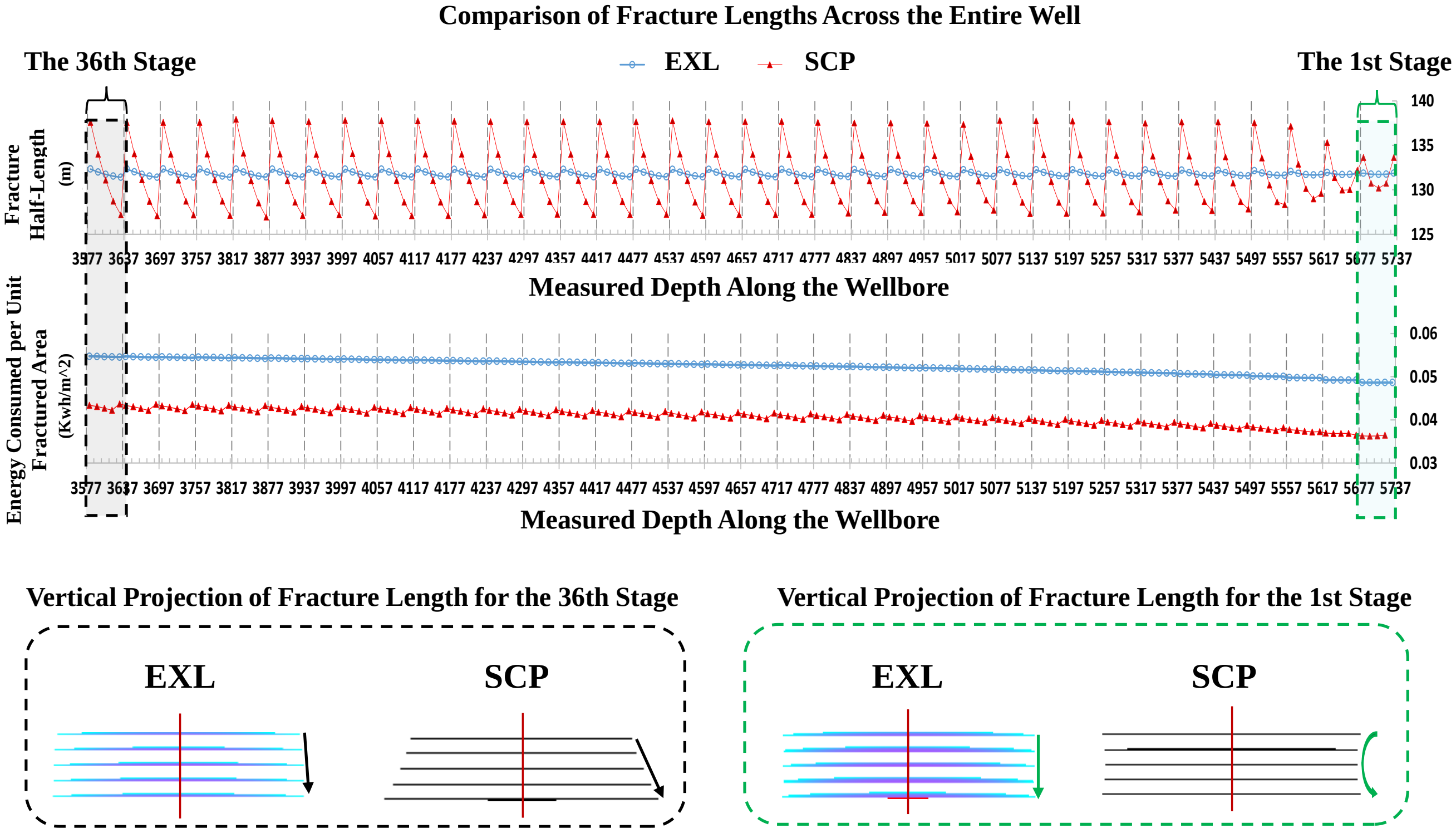

Figure 2: Comparison of Fracture Half-Length, Energy Consumption per Unit Fracture Area, and Fracture Expansion Between EXL and SCP

(The arrows in the enlarged view of the fracture length represent the trend of the fracture length within the chosen segment towards the toe direction along the wellbore)

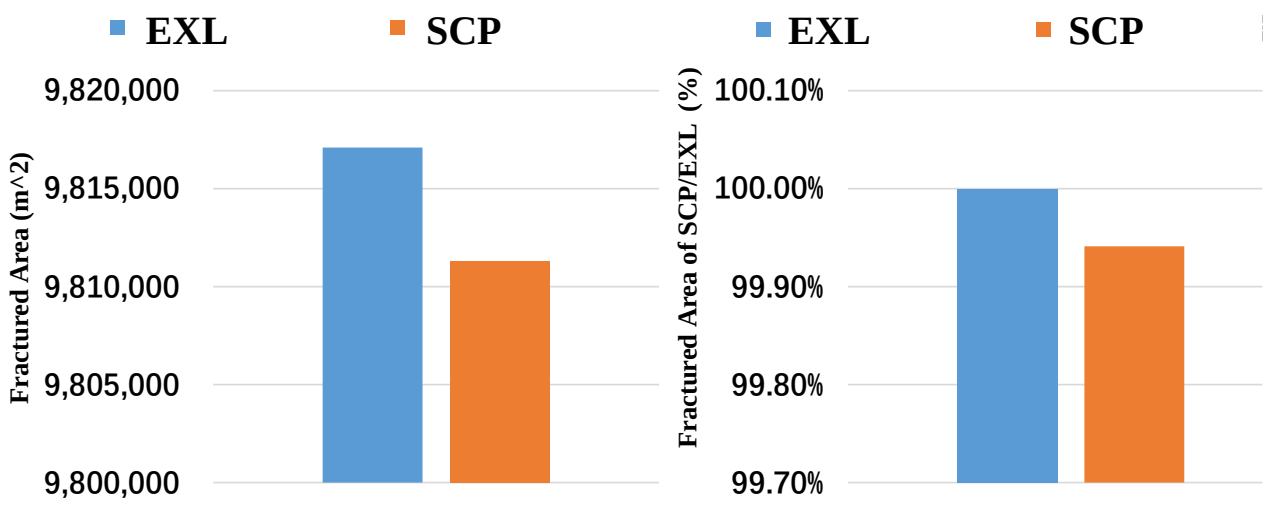

Figure 5 Comparison of Fractured Area

Table 1: Comparison of Perforation Parameters of EXL and SCP for a Homogeneous Reservoir in a Single Well Example

| Perforation Technique Type | Perforation Number/Cluster | Perforation Hole Diameter/(m) |
|---|---|---|
| EXL | 10 | 0.01 |
| G-SCP | 17 | 0.012 |

The simulation results show that no matter what kind of perforation technique is used, the fractures at both ends in the first three stages close to the toe are subject to less stress, allowing more fluid to enter and forming dominant fractures; the fractures in the middle are relatively suppressed. As stress accumulates, this phenomenon gradually changes in operation sequence: the length of the fracture is directly proportional to the distance from the previous operation segment. The Global Stress Coupled Perforation (G-SCP) technique, at the cost of losing 0.1% of fracture area, has a 22% lower energy consumption per unit fracture area than the Extreme Limited Entry (EXL) technique (see Figures 2 to 4). Therefore, for a single well in a homogeneous reservoir, the energy-saving effect of the Global Stress Coupled Perforation is indicated as significant.

3.2 Example of Stage Stress Coupled Perforation in Heterogeneous Reservoir for a Well Factory

During the zipper fracturing process in horizontal well factory, the heterogeneity of reservoir properties and rock mechanical properties, together with stress shadowing effects between clusters/segments/wells, make it difficult for fractures to expand uniformly. The stage by stage dynamic Stress Coupled Perforation (S-SCP) is more suitable for well group, changing with real-time fracturing procedure.

The Stage Stress Coupled technique (S-SCP) is applied in Fan Ye 1 Well Factory. The sweet spots in this well factory have an uneven spatial distribution, with the inter-well spacing of the four test wells varying from 350m to 800m (see Figure 5). Both Extreme Limited Entry technique (EXL) and Stage Stress Coupled Perforation technique (S-SCP) use conventional zipper fracturing sequences, alternating back and forth from one end of the well factory to the other for fracturing. That is, from Fan Ye 1-4 well to Fan Ye 1-7 well, then from Fan Ye 1-7 well to Fan Ye 1-4 well, in a cycle, until the last operation stage of the well factory is completed. The lateral length of each well is 1800m with 30 stages. Each stage is 60m long; five perforation clusters are symmetrically distributed in each segment, with a cluster spacing of 12.5m, and the clusters at both ends are 5m away from the bridge plug. The injected fluid is slick water, with a viscosity of 0.003Pa.s in 0.15$m^3/s$ rate. The total injection volume for each segment is 540$m^3$.

Due to the several orders' higher computation

Tabel 2 Perforation Parameters of S-SCP in First 3 Stages of FY1-6 Well

| Stage No | Upper Bound of the Perforation Clusters/(m) | Lower Bound of the Perforation Clusters/(m) | **"Perforation Length** /(m) | Number of Holes per Stage | Diameter of Perforation Holes/(m) |
|---|---|---|---|---|---|
| 1 | 5552.6 | 5553.4 | 0.8 | 40 | 0.006 |
| | 5540.1 | 5540.9 | 0.8 | | |
| | 5527.6 | 5528.4 | 0.8 | | |
| | 5515.1 | 5515.9 | 0.8 | | |
| | 5502.6 | 5503.4 | 0.8 | | |
| Plug | 5498 | | | | |
| 2 | 5492.7 | 5493.3 | 0.6 | 30 | 0.01 |
| | 5480.2 | 5480.8 | 0.6 | | |
| | 5467.7 | 5468.3 | 0.6 | | |
| | 5455.2 | 5455.8 | 0.6 | | |
| | 5442.7 | 5443.3 | 0.6 | | |
| Plug | 5438 | | | | |
| 3 | 5432.65 | 5433.35 | 0.7 | 35 | 0.008 |
| | 5420.15 | 5420.85 | 0.7 | | |
| | 5407.65 | 5408.35 | 0.7 | | |
| | 5395.15 | 5395.85 | 0.7 | | |
| | 5382.65 | 5383.35 | 0.7 | | |
| Plug | 5378 | | | | |
| 4 | 5372.65 | 5373.35 | 0.7 | 35 | 0.006 |
| | 5360.15 | 5360.85 | 0.7 | | |
| | 5347.65 | 5348.35 | 0.7 | | |
| | 5335.15 | 5335.85 | 0.7 | | |
| | 5322.65 | 5323.35 | 0.7 | | |
| Plug | 5318 | | | | |

efficiency than other numerical model with fully validated accuracy, multi-parameter coupling optimization of the accuracy [引用], the optimal combination of the perforation number and diameter was carried out, varying from 5 to 12 holes per cluster and the hole diameter varied between 6 and 10mm depending on the spatially-dynamic stress (Table 2 and Figure 5).

With similar fracture growth in euqal injection volumes, the S-SCP significantly reduces the energy consumption per unit fractured area compared to the Extreme Limited Entry (EXL). Taking well FY1-6 as an example, in the operation section from a measured depth of 4000m to 4238m, the fracture scale of the S-SCP was weaker than that of the EXL, with a difference of about 10m (see the gray area in Figure 6). Despite this, the fracture length of the S-SCP was relatively more uniform within four consecutive stages, especially in the 24th and 25th stages, where the S-SCP had 15m less variation in fracture length compared to EXL. The total fracture area and average energy consumption show that for well FY1-4, the total fracture area of S-SCP was 0.004% larger than that of EXL, meanwhile its energy consumption was reduced by 29%. This significant reduction in energy consumption with almost unchanged fracture area can also be observed in wells FY1-5, FY1-6, and FY1-7 (see Table1)

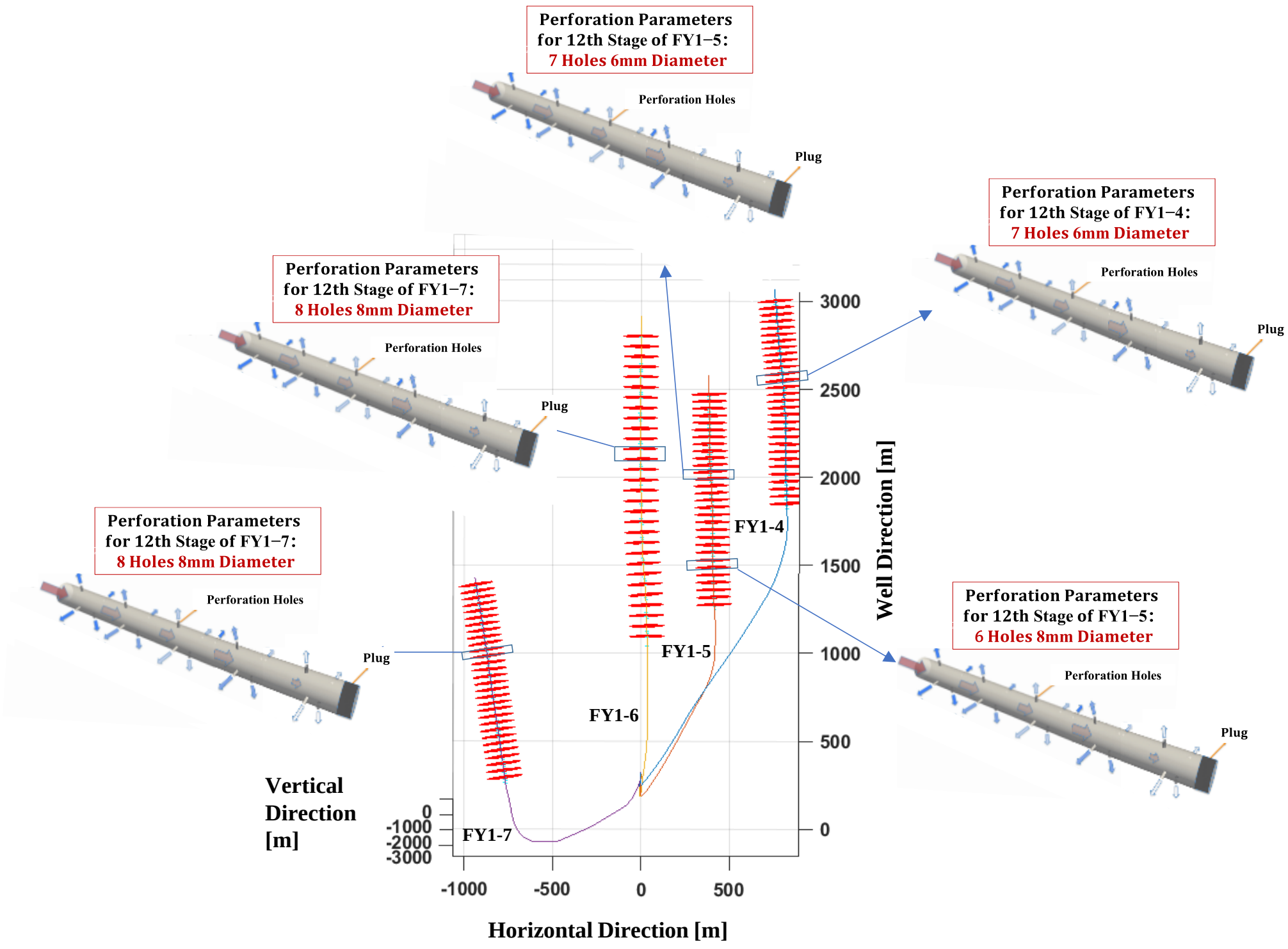

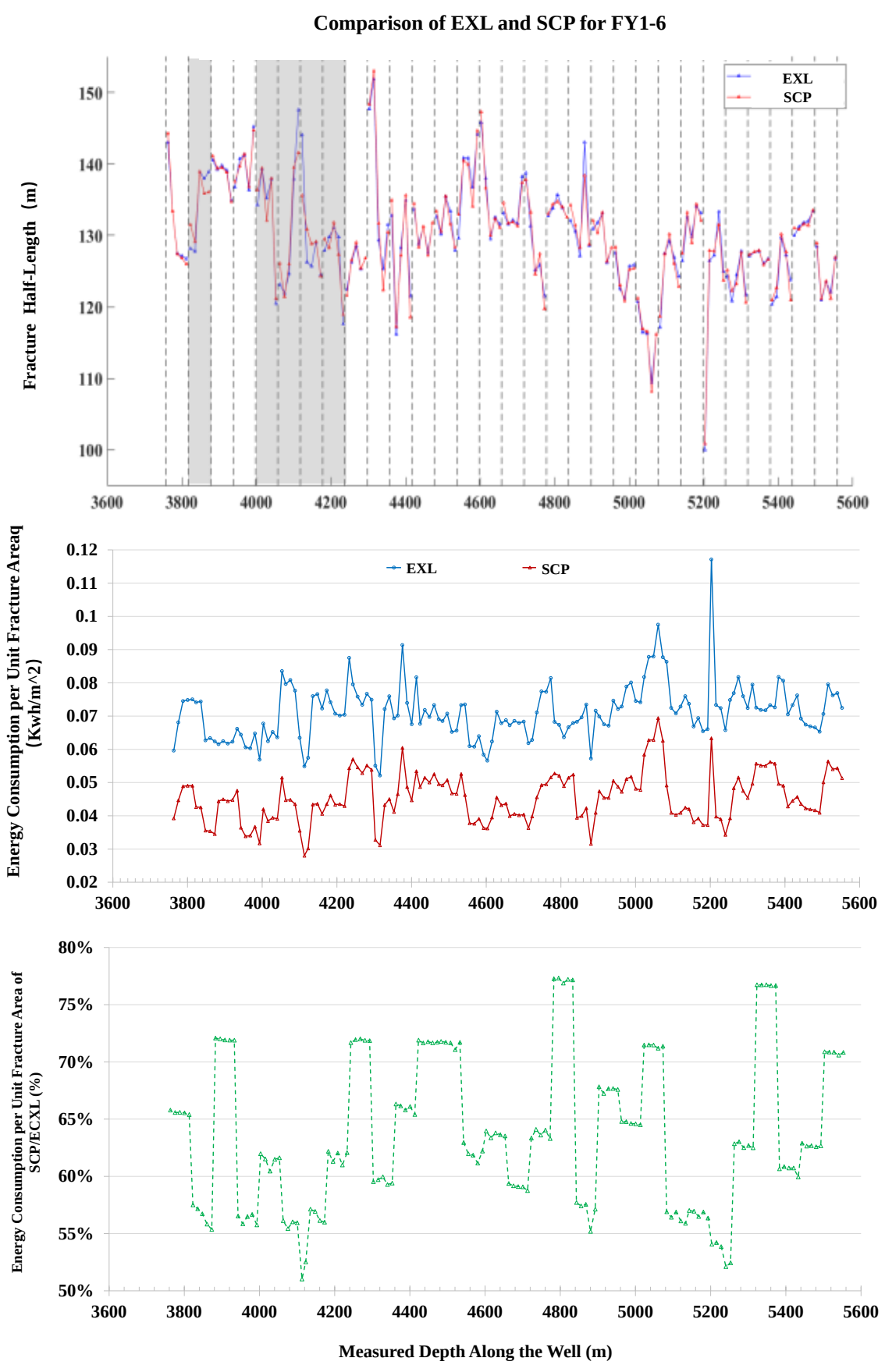

Figure 6: Comparison of Fracture Length and Energy Consumption per Unit Fractured Area in Fan Ye 1 Well Factory Using EXL and S-SCP (Dashed Line Represents the Position of the Plug)

Table 3 Comparison of Total Fracture Area and Corresponding Energy Consumption per Unit Fractured Area between EXL and SCP

| Comparison | Total Fractured Area[$m^2$] | | | Energy Consumption per Unit Fractured Area [$Kwh/m^2$] | | |
|---|---|---|---|---|---|---|
| Well No | EXL | S-SCP | Ratio(%) | EXL | S-SCP | Ratio(%) |
| FY**1-4** | 9077973 | 9078343 | 1.0004% | 0.0629 | 0.0451 | 71.586% |
| FY1-5 | 9685718 | 9682731 | 99.969% | 0.0601 | 0.0407 | 67.665% |
| FY1-6 | 7989279 | 7980892 | 99.895% | 0.0713 | 0.0455 | 63.783% |
| FY1-7 | 9005368 | 9001945 | 99.962% | 0.0678 | 0.0426 | 62.793% |

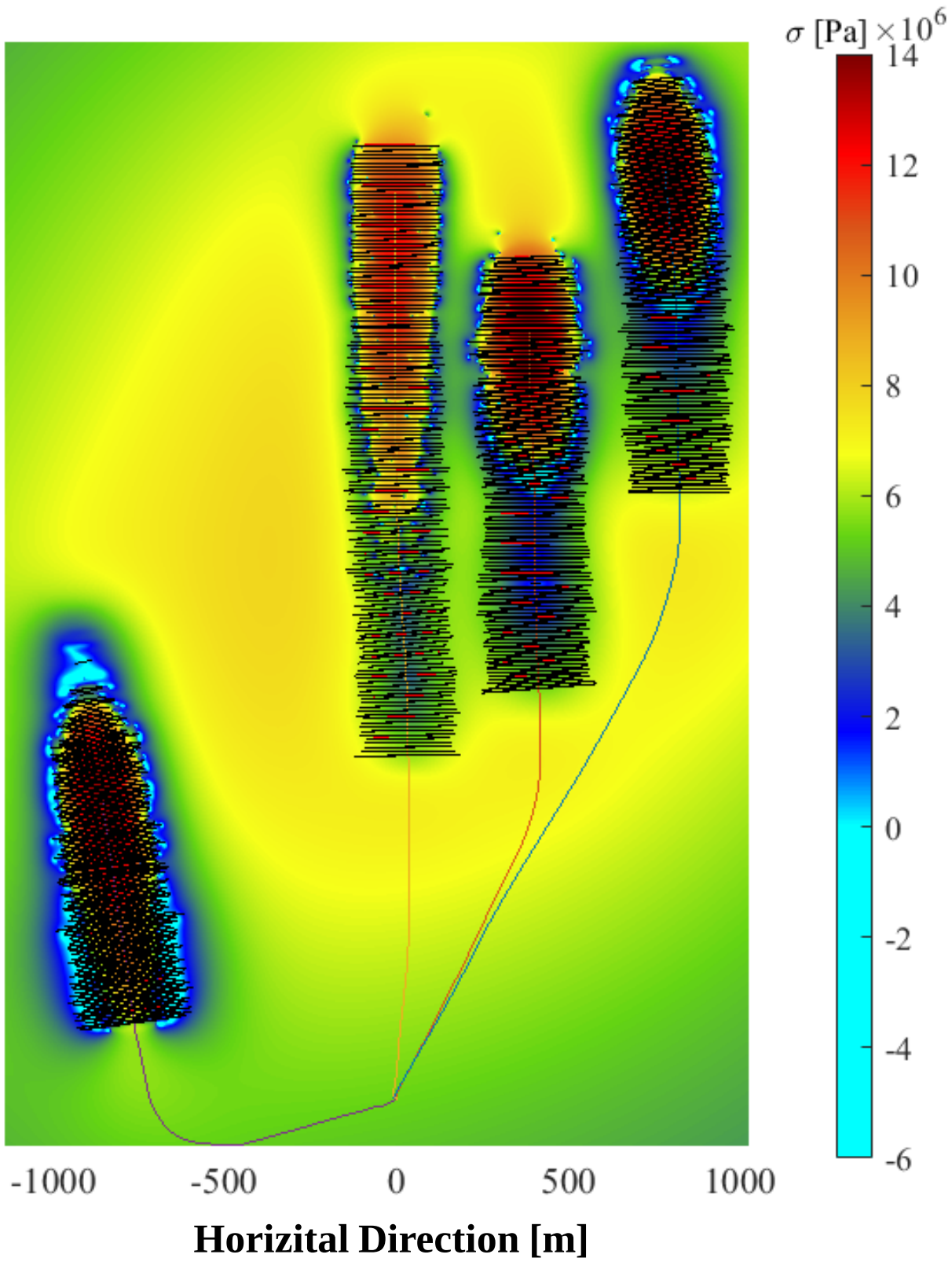

The distribution of formation-induced stress by the Stage Stress Coupled Perforation technique (S-SCP) indicates that stress generated by deformation is highly concentrated near the toe of the well. This is because the closer the well trajectory is to the target point (depth is approximately -3460m), the more horizontal the well trajectory becomes. The stress in the space is proportional to the distance from the center of the fracture. Therefore, at the horizontal slice depth for plotting, a compressive stress as high as 14MPa appears on the toe side (Figure 7). At the same time, the stress at the edge of the fracture is negative due to the tensile effect triggered by the initiation of the fracture at its end.

In order to more clearly demonstrate the impact of the two perforation techniques on stress distribution and also considering its' heterogeneity in three-dimension, a stress difference Δσ section is displayed every 100m starting from -3461.28m. The values on this section are obtained by subtracting the stress induced by the Extreme Limited Entry (EXL) from the stress stimulated by the Stage Stress Coupled Perforation (S-SCP).

Across all depths, it can be observed that the stress generated by the Stage Stress Coupled Perforation (S-SCP) are greater than that generated by the Extreme Limited Entry (EXL) (Figure 8). This is due to the different mechanisms of how these two perforation techniques drive fracture expansion.

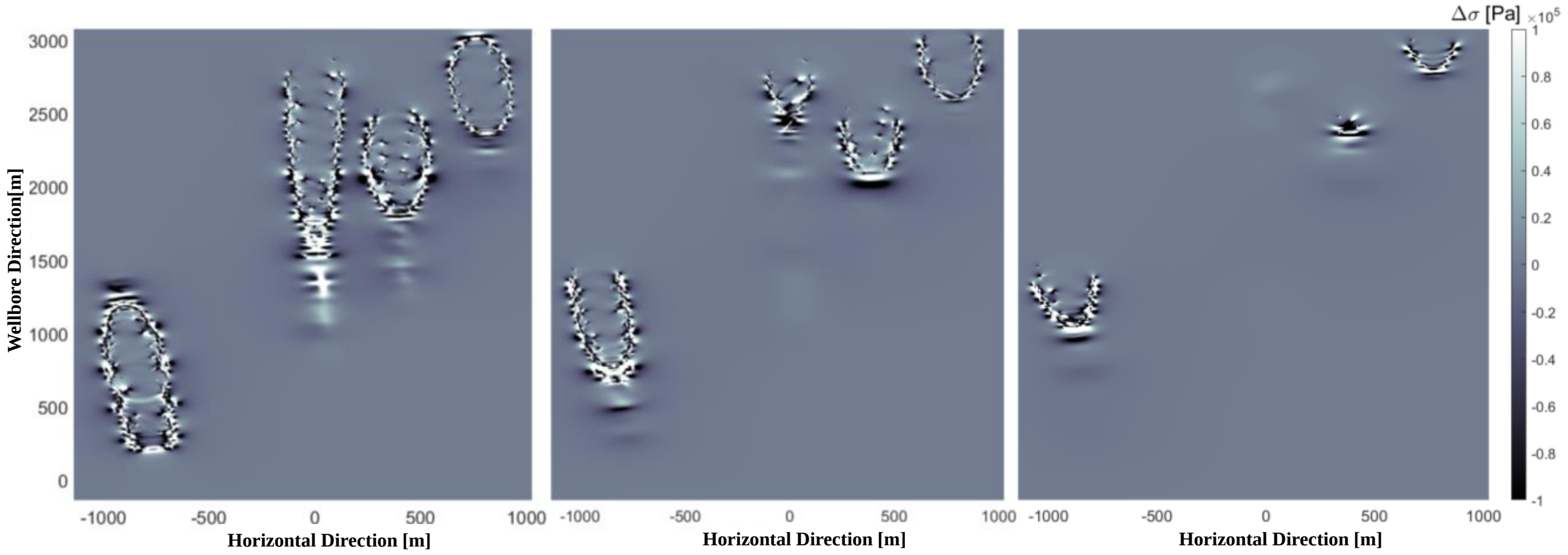

Figure 8: Horizontal Distribution of Induced Stress Difference in Fan Ye 1 Well Factory Using EXL and S-SCP

The EXL controls the stress distribution of the fracture system through a substantial pressure drop,

allowing the fractures to develop sufficiently under heterogeneous stress. In contrast, the S-SCP utilizes the elastic stress between fractures. By combining the tensile effect at the tip of the weakly developing fractures with the compressive effect on both sides of the superior fractures, it reduces the power needed to counteract the interaction stress (Equation 13). As a result, under the effect of the power balance Equation (14), it encourages more fluid to be allocated from the main fracture to the un-developed ones [29-32], thus achieving a more evenly developed fractures at a lower energy cost. Therefore, on the stress difference section, the stress difference between EXL and S-SCP is greater near the center and the edge of the fractures.

## 4 Conclusion

Hydraulic fracturing is a key technology widely used in the extraction of low-permeability/unconventional oil and gas. However, predicting the expansion of artificial hydraulic fractures is challenging due to the heterogeneity of subsurface reservoirs and the influence of operational parameters. The Extreme Limited Entry (EXL) drives fracture expansion by creating a significant pressure drop to dominate the stress distribution of the fracture system. However, as a cost, a substantial increase in pumping power appears, leading to increased greenhouse gas (GHG) emissions, negatively impacting the environment.

Stress-Coupled Perforation (SCP) reduces the power consumption required to counter other fracture expansions (Equation 13) by utilizing the tensile effects at the fracture tips and the compressive phenomena on sides. Under the boundary conditions of equal inlet pressure, it dynamically adjusts the inlet flow rate between the strong-developing fractures and the weakly-developing fractures to satisfy power balance, thus achieving unit fracture area at lower energy consumption. In the field trail of the Fan Ye 1 well factory, SCP reduced energy consumption by 37% while maintaining almost the same fracture area with the EXL.

In general, the wide application of Stress Coupled Perforation technique in hydraulic fracturing will have far-reaching implications for improving production efficiency and reducing environmental impact. In the future, with further enhancements in the computational efficiency of algorithms, the Clustered Stress Coupled Perforation (C-SCP) technique, with fine-tuning down to each cluster, may further resolve the contradiction between oil and gas resource extraction and the environment, providing more sustainable and environmentally friendly solutions for oil and natural gas development.

Acknowledgments: We appreciate the support from the National Natural Science Foundation of China and the US National Science Foundation. Special thanks to the guidance from my advisor Andrew Bunger (President of ARMA) and Academy of Engineering Emmanuel Detournay during the writing process.